\documentclass{eas}
\usepackage{graphicx}
\newcommand{\beq}{\begin{equation}}
\newcommand{\eeq}{\end{equation}}

\begin{document}

\title{Formation and tidal evolution of hot super-Earths in multiple planetary systems }
\author{Ji-Lin Zhou}\address{Department of Astronomy, Nanjing University, Nanjing 210093, China,zhoujl@nju.edu.cn}
\begin{abstract}
Hot super-Earths are exoplanets with  masses  $\le 10 M_\oplus$ and
orbital periods $\le 20$ days. Around 8 hot super-Earths have been discovered in the neighborhood of  solar system.
In this lecture, we review the mechanisms for the formation of hot super-Earths,  dynamical effects that play important roles in sculpting
the architecture of the multiple planetary systems. Two example systems (HD 40307 and GJ 436) are presented to show the formation and evolution of hot
super-Earths or Neptunes.
\end{abstract}
\maketitle

\runningtitle{Formation and evolution of hot super-Earths}

\section{Introduction}

More than 330 exoplanets have been detected in the neighborhood of solar system by various techniques, especially by
radial velocity measurements\footnote{http://exoplanet.eu/}.
Among them, hot super-Earths(HSEs hereafter) are characterized with  masses  $\le 10 M_\oplus$(Earth mass) and
orbit periods $\le 20$ days. The  mass upper limit is set based on the critical mass ($\sim 10 M_\oplus$) in the core-accretion  scenario of
giant planet formation,  above which efficient gas accretion will set in(\cite{pol96}). The period upper limit ($20$ days, or $\sim 0.15$ AU)
is set with somewhat arbitrary, where the
temperature of a solid body due to the radiation of a solar-mass star is $\sim 730$K, and the tidal circularization timescale of
a $10M_\oplus$ solid planet is $\sim 13$ Gyrs, thus tidal dissipation is effective for solid planets inside this orbit.

To date,  only 8 HSEs are detected, with 13 hot Neptunes  (HNs hereafter) which have similar orbital periods ($<20 $days) but larger masses.
These 21 HSEs and HNs are distributed in 15 planetary systems,  among them
there are 5 single-planet systems up to the present observations: GJ 436, GJ 674, HD 219828, HAT-P-11, HD 285968, HD4308; the others are in
multiple planet systems(Fig.1.). Due to the limit number of HSEs and HNs,  a statistical study of the orbital parameters for HSEs  is still unreliable.

\begin{figure}[!htbp]
\vspace*{0cm}
\begin{center}
\includegraphics[width=5.0in]{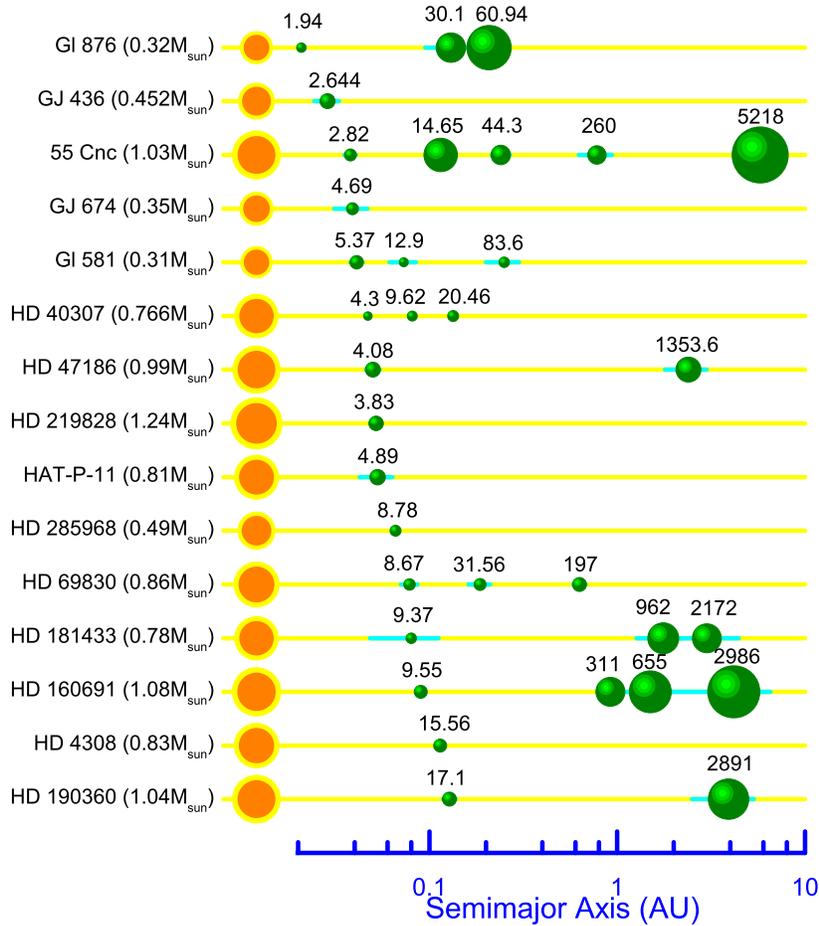}
\vspace*{0cm}
 \caption{\small Diagram of semimajor axes and masses for the 15 known hot super-Earth and Neptune systems.
 The diameters depicted for planets and stars are
proportional to the cube root of the planetary $M \sin i$ and stellar mass, respectively.
 The periapse to apoapse excursion is shown by a horizontal line intersecting the planet. Labels around the planets
 are their orbital periods in units of days. Data are from http://exoplanet.eu/.}
   \label{fig1}
\end{center}
\end{figure}

One interesting problem is the formation of HSEs. According to the core-accretion model of planet formation(\cite{Saf69,pol96}),
 planetesimals coagulate via runaway growth and become protoplanetary embryos through runaway and oligarchic growth(\cite{KI96,KI02}).
 When one adopts the empirical minimum
mass solar nebula model (hereafter MMSN) ,
 the surface density of gas and the heavy elements are expressed as(\cite{hay81,idalin04}),
\begin{equation}
 \Sigma_{g}=2.4\times 10^3 f_g  (\frac{a}{ 1 {\rm AU}}  )^{-3/2}~{\rm g~cm}^{-2},
 \label{sigg}
\end{equation}
and
\beq
\Sigma_{d}=10 f_d \gamma_{\rm ice}   (\frac{a}{ 1 {\rm AU}} )^{-3/2}~{\rm g~cm}^{-2},
\label{sigd}
\eeq
respectively, where $a$ is the semi-major axis, $f_g$ and $f_d$ are enhancement factors,
$\gamma_{\rm ice}$ is the volatile enhancement with a value of  4.2 or 1 for exterior or interior to the snow line
(where temperature is 170K due to the stellar radiation), respectively.
In such a disk,  the  growth timescale of a planetary embryo with mass $M$ is estimated as(\cite{KI02}),
\begin{equation}
\tau_{\rm growth} \simeq 0.12 \gamma_{\rm ice}^{-1}
f_d^{-1} f_g^{-2/5} \left( {a \over 1 {\rm AU} } \right)^{27/10}
\left( {M \over M_\oplus} \right) ^{1/3} \left( {M_\ast \over
M_\odot} \right)^{-1/6} {\rm Myr},
\label{eq:tgrowth}
\end{equation}
where $M_*$ is the stellar mass. In the ideal situation that planetesimals do not undergo significant orbital decay,
the growth of embryos are stalled with isolation masses(\cite{idalin04}),
\begin{equation}
M_{\rm iso} = 0.16 f_d^{3/2} \gamma_{\rm ice} ^{3/2}
(\frac{\Delta_{\rm fz}}{10 R_h})^{3/2} (\frac{a}{\rm AU})^{3/4}
(\frac{M_*}{M_\odot})^{-1/2} M_\oplus,
\label{miso}
\end{equation}
 when they cleaned the planetesimals within their feeding-zone of width
$\Delta_{\rm fz} \sim 10-12 R_h$, where $R_h = (2M/ 3 M_\ast )^{1/3}
a$  are their mutual Hill's radii.
The orbital crossing timescale for a swam of isolated embryos with equal mass ratios
$\mu=M_{\rm iso}/M_*$ is fitted by numerical simulations (\cite{zls07})
\beq
 \log (\frac{T_c}{\rm P_m}) =A + B \log(\frac{\Delta_{\rm fz}}{2.3R_h}),
 \label{ff}
\eeq
where
\beq
A=-2-0.27\log\mu, B=18.7+ 1.1\log\mu,
\label{ab}
\eeq
and $P_m$ is the period of the middle planet ($\sim 10$ days for HSE).
For a moderate disk with $f_d=2$ and feeding zone width of $12 R_h$,
 the mass of an isolated embryo at $10$ day's orbit is $M_{\rm iso}=0.08M_\oplus$ with a very short
 growth time (within 100 years), however, the orbital crossing timescale for such a swarm of embryos is $\sim 7$ Gyr by Eq.(\ref{ff}),
 which  it is quite stable unless extra perturbations excite their eccentricities. Considering the observed HSEs,
 the interesting question is: how do they form from a swarm of isolated embryos?

\section{Formation Scenarios of HSE}

Several scenarios for the formation of  HSEs  are summarized in Raymond et al. (2008).  
  There are three major sources of HSEs:
(i) embryos shepherded into the mean motion resonances by inward migration of  gas giants in
outside orbits(\cite{zhou05,FN05}),
(ii) embryos  shepherded  by the secular resonances between gas giants
during the depletion of gas disk(\cite{zhou05,nag05}). In these two mechanisms, the eccentricities of the embryos  are excited  in resonances, which
results in their merge and growth to HSEs.
(iii) inward type I migration of super-Earth protoplanets to hot orbits.
As we can see from eq.(\ref{miso}),
in situ formation of HSEs requires a large solid mass ($f_d\ge 10 $),  which is unlikely unless type I migration of
embryos are effective.
In fact, for a specific system, all these effects (type I migration, gas giant perturbation or disk depletion, tidal dissipation, etc.)
 may account for making  the architecture of a planet system with HSEs.
 We discuss several major effects in the following text.

\subsection{Type I migration}

One of the major obstacle for the core-accretion scenario is
the fast migration of Earth mass embryos. For such  a planet embedded in a geometrically thin
protoplanetary disk, angular momentum exchanges between the planet
and nearby  gas disk will cause a net momentum lose on the planet, which results in a so called type I
migration of the planet with a timescale(\cite{GT79,Ward97,Tan02}),
\begin{equation}
\tau_{\rm I-mig} \simeq \frac{1}{C_1 (1.1 \beta-2.7)} \left(\frac{M_*}
{M}\right) \left(\frac{M_*}{\Sigma_g a^2}\right)\left( \frac{H}{a}
\right)^2 \left(\frac{P_K} {2 \pi}\right)
\label{tauI}
\end{equation}
where negative/positive values of $\tau_{\rm I-mig}$ corresponds
inward/outward migration respectively, $M$, $P_K$ and
$H$ are the embryo's mass, Keplerian period and
thickness of the protostellar disk, respectively, $\beta \equiv
\partial {\rm ln} \Sigma_g/ \partial {\rm ln} a$ and $C_1 (\sim
0.03-0.1)$ is a reduction factor.
According to Eq. (\ref{tauI}), the migration timescale is
$0.05$ Million years for a Earth mass
planet at 1AU. 
Such a fast migration will evacuate the embryos so that no gas giant will form through core-accretion scenario,
unless the migration speed is at least an order of magnitude smaller(\cite{ali05,idalin08}).
 How to  reduce the speed of type I migration is a key
issue of present planet formation theory.

Recent hydrodynamical simulation indicates that
type I inward migration can be stopped near a boundary of
a density maximum(\cite{mas06}).  During the orbital decay of a protoplanet,
the exchange of angular momentum between it and the fluid elements
that perform U-turn at the end of the horseshoe streamlines
generates a corotation torque on the
protoplanet(\cite{ward91,masset02}).  The quantity of
corotation torque, $\Gamma_c \propto \Sigma_g d\ln
(\Sigma_g/B)/d\ln a$, depends on the local gradient of the disk
surface,where $B$ is the Oort constant and $B  \sim a^{-3/2}$ in a
near Keplerian motion.
At the surface density increment, the corotation torque of the
disk  is  positive. 
 Simulations
indicate that a disk-surface-density jump of about $50 \%$ over 3-5
disk thicknesses suffices to cancel out the negative Lindblad
torque that generates the type I migration, leading the density maximum a ``trap" of protoplanet(\cite{mas06}).

Several locations of the protoplanetary disk can serve as the density maximums, e.g., the
{\em  inner disk cavitary} due to the stellar magnetic field.
Around the corotation radius, the stellar magnetic toque acts to
extract angular momentum from the disk and spins down the disk
material. 
At the location where the stellar magnetic field completely dominates
over disk internal stresses, sub-Keplerian rotation leads to a
free-fall of disc material on to the surface of the star
 in a funnel flow along magnetic-field lines, results in an inner disk
 truncation(\cite{Kon91}).
 The  maximum distance of disk  truncation is estimated at $9.1$
 stellar radii. Considering  the radius of protostar is generally 2-3 times
  larger than their counterpart in main sequence, the inner disk truncation would occur at $\sim 0.1$AU.
Once the embryo has spiral in the inner disk cavitary, type I
migration is greatly reduced according to equation (\ref{tauI}),
thus migration is effectively stopped.

Another place is the {\em Boundary of MRI active-dead zones}. 
During the classical T Tauri stars (CTTS) phase,  magnetorotational
 instability (MRI) is  effective so that the gas disk has a layered structure as a sandwich: inside the boundary,
 the protostellar disk is thermally ionized while outside this boundary, only the
 surface layer (with a thickness  $\cong 100 $g cm$^{-2}$)
is ionized by stellar X-rays and diffuse cosmic rays, leaving the
central part of the disk a highly neutral and inactive
``deadzone"(\cite{BH91,gam96}). Near the boundary of active and dead
zone, a positive density gradient is expected(\cite{KL07,KL09}).
In fact, adopt the  ad hoc $\alpha$-prescription  for  the protoplanetary gas disk(\cite{SS73}), the effective viscosity  of the disk is written as
$\nu=\alpha c_s h$, where $c_s$ and $h=c_s/\Omega_K$ are the
sound speed of midplane and the isothermal density scale
height,respectively. The magnitude of $\alpha$ increases about two
times  from magnetic saturated regions ($\sim 0.006$) to  active
ones($\sim 0.018$). Assuming a constant mass accretion rate
($\dot{M}_g=3\pi \nu \Sigma_g$) across the disk,
the variation of viscosity $\nu$ indicates $\Sigma_g$ could increase
two times from MRI dead zone to active zone.
The sharp increase of column density across the boundary  helps to halt the embryos  under type I migration.

According to the above discussions,  the inward migration of type I
 can be halted at some specific location of the disk before gas depletion.
  The observation of HSEs, e.g.,  HD 40307 system(section 3.1),  gives a clear
evidence for the migration-and-halt history.

\subsection{Perturbations from planet companions}

As we have shown in  Eq.(\ref{miso}), embryos outside the snow line tend to have large isolation masses.
When their masses exceed the critical mass ($\sim 10M_\oplus$) so that efficient gas accretion will set in, they will grow to gas giants in a timescale
of million years, leaving a group of
isolated embryos with masses $\sim 0.1 M_\oplus$ inside the snow line.
 During and after the formation of gas giants, their evolution
will greatly affect the subsequent formation of super-Earths. There are two major types of perturbations to these embryos:

{\em Shepherding in Mean Motion Resonance (MMR)}. During the type II migration of gas giants,
the locations of inner mean motion resonances (mainly 2:1 MMR) with the gas giants will sweep through inner disk,
trap and shepherd the embryos, excite their eccentricity, which results in the merge of isolated embryos and formation of HSEs(e.g., \cite{zhou05}).
The resonance trap and shepherding is efficient as long as the timescale of type II migration(\cite{idalin04}),
\begin{equation}
\tau_{\rm II}=0.8 {\rm Myr}~
f_g^{-1} (\frac{M_p}{M_{\rm
J}})(\frac{M_{\odot}}{M_*})(\frac{\alpha}{10^{-4}})^{-1}(\frac{a_p}{\rm
1AU})^{1/2},
\end{equation}
is much longer than the libration period (typically less than hundreds of years inside the snow line),
 where $M_p,a_p$ are the mass and semi-major axis of the gas giant.
Prior to severe gas depletion, the eccentricities of the embryos
are damped on a timescale(\cite{ward93}),
 \beq
\tau_{\rm e, damp} \simeq 300
f_g ^{-1}  (\frac{M}{M_\oplus})^{-1} (\frac{a}{1 {\rm AU}})^2 {\rm yr}.
 \eeq
The eccentricity damp of embryos enhances the trap of 2:1 MMR, as the resonance region for circular orbits is relatively wider.
However, when  type I migration of  embryos are taken into consideration, whether they will stop at 2:1 MMR or other MMRs with higher orders depends
on the migration speed of embryos, as we will see in section 3.1 for HD 40307 system.

After the embryos become super-Earths, their orbits may perturb each other, or they can accrete gas to become
 giant planets, if there exists  efficient gas in the disk.  The embryos themselves may trap into resonance
through type I migration.  Recently simulations indicate that resonance trap between embryos are common during the migration of
embryos (\cite{tp07,FN07,Mor08,lin09}).

{\em Secular perturbations} between two planets (or embryos) in non-resonance orbits
exchange their angular momentums, modulate their
eccentricities, leaving their semi-major axes almost unchanged.
Under the perturbation of an outer planet ($M_2$) with semi-major axes $a_2$ and eccentricity $e_2$,
the maximum eccentricity $(e_1) $ of the inner planet ($M_1$) that can achieve  from an initial circular orbit $a_1$ is,
\begin{equation}
e_{\rm 1max}=\frac{5}{2}  e_2\varepsilon_2^{-2}(\frac{a_1}{a_2})
\left|1-\varepsilon_2^{-1}\sqrt{\frac{a_1}{a_2}} \left(
\frac{M_1}{M_2}\right)+\gamma\varepsilon_2^3\right|^{-1}.
\label{e1max}
\end{equation}
where $\varepsilon_2=\sqrt{1-e_2^2}$, $\gamma$ 
 is the ratio of
general relativity to companion perturbation on periapsis
precession of $M_1$(\cite{mar07}).
This equation can be used to locate an approximate region of the
planet companion in either nearby or distance orbits,
 while general three-body simulations should be performed to give a precise
location.

\subsection{Tidal evolution}
A close-in planet produces tidal bulges on the stellar surface,
causing energy dissipation on the star and angular momentum
exchanges between the stellar spin and planetary orbital motion.
 Meanwhile the star also generates tidal dissipation on the planet, resulting in an
 eccentricity damping and  orbital decay.
 For close-in planets with tidal dissipation
factor $Q' \le 10^6$, dissipation in planets dominates.
The timescale of orbital
circularization ($\tau_{\rm circ}=e/\dot{e}$) induced by planetary
tidal dissipation is given as(\cite{ML04,ZL08}),
\begin{equation}
\tau_{\rm circ}=2.4\times 10^7  Q'_1
  (\frac{a}{\rm 0.1AU})^{\frac{13}{2}}
  (\frac{M_*}{M_\odot})^{-\frac{3}{2}}
  (\frac{M}{M_\oplus})^{-\frac{2}{3}}
   (\frac{\rho}{\rm 3 g~cm^{-3}})^{\frac{5}{3}}~ {\rm yr},
\end{equation}
where $\rho$ is the density of the planet. The associate timescale of orbital decay ($\tau_{\rm decay}=
a/\dot{a} $) in elliptical orbits is \beq \tau_{\rm decay}
\approx \frac{1-e^2}{2e^2} \tau_{\rm circ}.
\eeq
The orbital decay of a HSE under tidal dissipation determines its final location.

\section{Examples systems}
\subsection{HD 40307 system}

\begin{figure}[!htbp]
\vspace*{3.5cm}
\begin{center}
\hspace*{2cm} \includegraphics[width=5.0in]{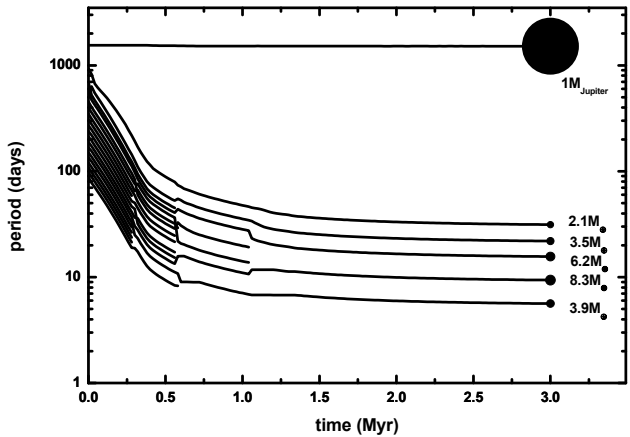}
\vspace*{-6.5cm}
 \caption{\small Evolution of embryos in one run of simulation for HD 40307 system, where 20 embryos  with initial
 isolation mass of Eq.(\ref{miso}) with $f_d=6$ are put in inner orbits.  A Jupiter mass planet  is
 put at 2.4 AU without type II migration.  Embryos undergo type I migration during evolution, with
 a reduction factor $C_1=0.3$ in Eq. (\ref{tauI}). }
   \label{fig2}
\end{center}
\end{figure}

The recent observed extra-solar planetary system around star HD40307
hosts three super-Earths with masses 4.2$M_\oplus$,
6.9$M_\oplus$, 9.2$M_\oplus$ in orbits of 4.3days, 9.6 days
and 20.5 days, respectively(\cite{Mayor08}). This configuration is very close to
 the Laplace resonance (with mean motion locking to $4:2:1$) among the three Galilean Satellites,
  Io-Europa-Ganymede,  which is the only known example with such a configuration.

Lin et al. (2009) studied the formation of the HSEs in HD 40307 system.
In situ formation of these  planets  requires a high density
so that the disk must have been enhanced in
refractory solids with $f_d > 25$, which  is difficult to
achieve in a gravitationally stable disk around a $0.8 M_\odot$ and
metal-deficient star. A compact system of embryos with smaller $M_{\rm
iso}$ can formed under less extreme conditions and coagulate after the
gas depletion. Fig.2 shows a typical run of N-body simulation with a Hermite-type code(\cite{aarseth}).
In this run  a  disk model of (\ref{sigg}) and (\ref{sigd}) with $f_g=1$  and $f_d=6$ are employed.
In the inner region of disk, the column density increase near the
boundary of MRI active-dead zone is considered(\cite{KL07}).
Reasonable  parameters suitable for CTT stars  estimate the boundary at around 20 days.
We also assume that the disk has a
exponential decay with a timescale of 1 Myr(\cite{Hai01}).
We put initially 20 embryos with isolation masses from (\ref{miso}) and mutual separation of 7 Hill's radii,
and one Jupiter mass giant planet located at 2.4 AU. After the evolution of 3 Myrs,
the final system compose 5 super-Earths, with much smaller mutual separation. This simulation indicates that in situ formation
of the three planets in HD 40307 system is unlikely.

Thus the most plausible scenario for their formation is (\cite{lin09}):
 (i) formation of three planets  in outside orbits; (ii) type I migration of the planets and the halt of
 migration  of $m_1$ due to the density profile enhancement near the boundary of MRI active-dead zone($a_{\rm mag}$);  the ongoing
migration of $m_2$ and $m_3$ leads a consecutive trapping of 2:1
resonance with $m_1$ and $m_2$, respectively,  which results in a $4:2:1$ configuration among the three planets, (iii)
tidal evolution of three super-Earths during and after gas disk
depletion to the present locations.
As a typical run, Fig.3a  shows that after the migration of  $m_1$ is slowed down and stalled at $a_{\rm mag}$ at $T=0.2$ Myr,
 the approaching $m_2$ enters into a 2:1 MMR at $T=$ 0.3
Myr. Subsequently planet 3 is captured into $m_2$'s 2:1 MMR at $T=0.6$
Myr, which results in a $4:2:1$ resonance.
 However, if we increase the speed of type I migration by setting $C_1=0.3$ in Eq. (\ref{tauI}),
then the three planets will fall in a $6:3:2$ resonance.
 A standard speed of type I migration ($C_1=1$) will reach a final configuration
of $9:6:4$. These results indicate clearly that the
trap of mean motion resonance depends on the timescale of
migration.

Fig.3b shows the subsequent tidal  evolution of three planets
initially  from $4:2:1$ resonance.  Due to the tidal
evolution and resonance interactions, the  planets  move inward
 while keeping the resonance configuration unchanged. This
requires $\Delta a/a$ being  the same for three planets, which is
most difficult to follow for the outmost planet since its tidal
dissipation is lowest. The configuration is broken at time $T \sim
0.7$Myr when $m_3$ first leaves the resonance.
The final configurations  have periods of
$4.35$days, $9.53$days, $20.15$days, eccentricities of 0.016,
0.020,0.027 for $m_1$,$m_2$,$m_3$, respectively. The relative errors
of final periods compared with observational data of HD 40307 system are $0.9\%, 0.9\%
1.5\%$, relatively.
 Due to the linear dependence
of tidal force on $Q'$, the final state of circular orbits is achieved at time
$T \sim 5 Q'$Myr.
 To compare with the observed circular orbits, this gives a restriction of
$Q' < 2000$ supposing an age of 10 Gyr for the star HD40307, indicating these HSEs are most likely rocky planets.

\begin{figure}[!htbp]
\vspace*{4.5cm}
\begin{center}
\includegraphics[width=7.0in]{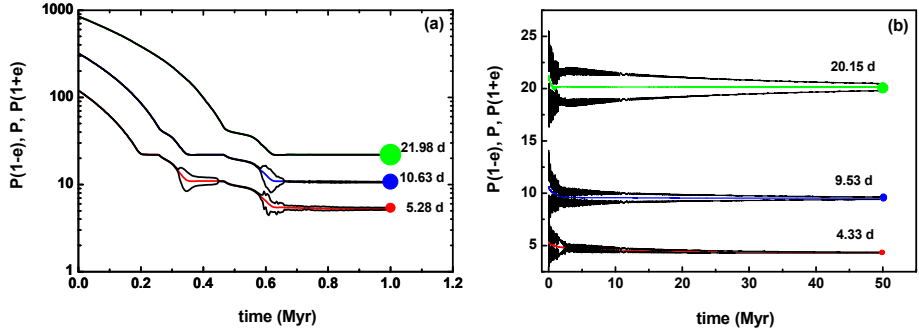}
\vspace*{-13cm}
 \caption{Evolution of three super-Earths system around HD 40307 with reduced
Type I migration speed, $C_1=0.1$ in Eq. \ref{tauI}. Three planets initially locate
in  circular orbits with periods of $120,320,850$days.
 Panel (a) presents the evolution of $P(1+e), P,
P(1-e)$ of each planet, where $P$ is the orbit period and $e$ is the
eccentricity.  Panel (b) shows the tidal evolution of three
super-Earths systems  with  an illustrative $Q'=10$. Initially  three
planets are set  in $4:2:1$ MMR (with periods of $5.3$days, $10.6$days, $21.2$days,
respectively),  with initial
eccentricities  $0.20$ for all the three planets.  From Lin et al. (2009).  }
   \label{fig3}
\end{center}
\end{figure}

\subsection{GJ 436 system}

 GJ 436b is a Neptune-size planet with 23.2 Earth masses in an
elliptical orbit of period 2.64 days and eccentricity 0.16 (\cite{But04,man07}, etc).  With
a  typical tidal dissipation factor ($Q' \sim 10^6$) as that of a
giant planet with convective envelope,
 its orbital circularization timescale under internal tidal dissipation is around $1$ Gyr,
 at least two times less than the stellar age ($>3$~Gyr).
Considering that radial velocities of GJ 436 reveal a long-term
trend, Maness et al.(2007) 
proposed the presence of a long-period ($\sim
25$ yr) planet companion with mass $\sim 0.27 M_J$ (Jupiter mass)
in an eccentric orbit ($e\sim 0.2$). Recently, 
Ribas et al.(2008) suggested that the observed radial velocities of the system
are consistent with an additional small, super-Earth planet in the
outer 2:1 mean-motion resonance with GJ 436b.
 More recent inspection of transit data implies
that GJ 436b is perturbed by another planet with mass $\le 12
M_\oplus$ in a non-resonant orbit of $\sim 12$ days (\cite{cou08}).

In Tong \& Zhou (2009), we  investigated extensively  the possibility of the eccentricity excitation of GJ 436b
by a companion, assuming  the companion is either in a nearby/distant orbit, or in MMR with GJ 436b.
Fig.4 shows the maximum eccentricity in parameter space that can
  be excited by the companion in nearby orbits.
  We find that,
although the eccentricity of GJ 436b can be
excited to 0.16 with a broad range of companion mass (above few
Earth-masses), the maintain of the eccentricity to 0.16 under tidal dissipation is impossible.
 In fact,  as the orbital decay
time is  short ($\sim 20 $Gyr) for GJ 436b at the present location, significant orbital decay $(\sim
25\%)$ is expect so that GJ 436b would be in a much closer orbit,
and the eccentricity of GJ 436b would be damped within the stellar
age.  Distance companions can {\em not}
excite and maintain the significant eccentricity of GJ 436b unless they are in highly eccentric orbits.

Based on the extensive investigations, we think
the high eccentricity of GJ 436b can not be maintained by a companion  in
either nearby or distance orbits through secular perturbation or
mean motion resonances. These results do {\em not} rule out the possible existence of
  planet companions in nearby/distance orbits, although
they are not able to maintain the eccentricity of GJ 436b.
Thus the maintaining of its eccentricity
remains a challenge problem, unless GJ 436b has a extremely high
dissipation factor ($Q'> 6 \times 10^6$).

\begin{figure}[!htbp]
\vspace*{-2.0 cm}
\includegraphics[scale=1]{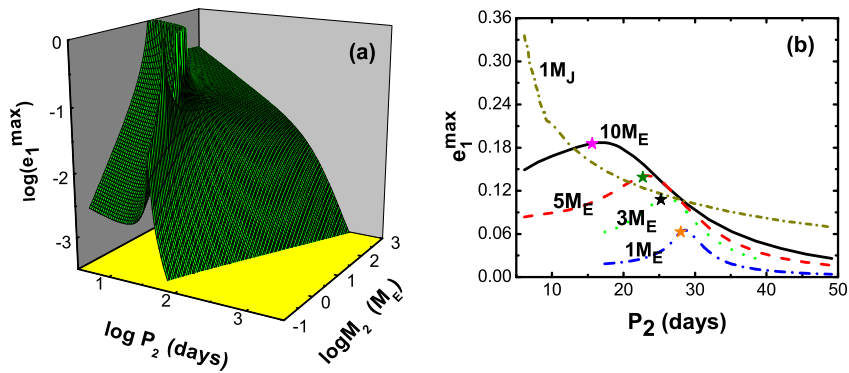}
\vspace*{-2.0 cm}
 \caption{\small Maximum eccentricity of  GJ 436b  that can be excited at the present location
   by a planet companion  in outside orbit with initial eccentricity $e_{2}=0.2$. Panel (a): 3-D plot of
   $e_{\rm 1max}$ in the plane of  $P_2-M_2$ from theoretical Eq. (\ref{e1max}).
   Panel (b): $e_{\rm 1max}$ obtained from numerical simulations of a general three-body model.
    The asterisks denote the singularities from Eq. (\ref{e1max}). From Tong \& Zhou (2009).}
   \label{fig4}
\end{figure}

\section{Conclusions}

As more and more hot super-Earths (HSEs) are detected recently,  with few  multiple
HSE systems, their formation story is not fulled understood yet.  For example, there exist several mysteries.

{\em Retention of planetesimals and embryos } under gas drag and disk tidal decay remains the major challenge.
Although mechanisms have been proposed to stop the inward migration of Earth mass planet near the inner cavitary of disk or
near the boundary of MRI active-dead zone, the ubiquity of
super-Earth planets at many other locations of the system merits  explanation.
If indeed they can only be stopped in the boundary MRI active-dead zone,
or the inner cavitary of disk, observation of HSEs can help to estimated such locations in the CTTS stage of protostars.
During the later stage of planet formation,
UVO photoevaporation from the host star or cosmic rays may severely deplete the gas disk,  which will reduce  the surface density
of the disk by an order of magnitude at $\sim 1$ Myr(\cite{alex06}). Thus the migration speed is an order of magnitude
lower than that estimate from the linear theory,  embryos  formed in the later stage could survive at more distant orbits.

{\em Final configuration of Earth-mass planets}. Most simulations of HSE formation indicates the presence of several super-Earths formed within
mutual mean motion resonance (MMR) under type I migration. However,  only HD 40307 system is
observed to be  near MMR. Tidal evolution will play part of the roles by removing such configurations,
 leaving the multiple planets in near resonance instead of the exact resonances. On the other hand, system with Earth mass planets
 far from resonance configurations are also present, like our solar system. Why type I migration seems to be not effective in such systems
 requires further investigations.

{\bf Acknowledgements} I would like thank the organizers of the conference for their kind invitation.
 This work is supported by NSFC (10833001,10778603), National Basic Research Program of China (2007CB814800).



\begin{thebibliography}{99}
\bibitem[Aarseth 2003]{aarseth}
 Aarseth, S. J. 2003, {\it Gravitational $N$-Body Simulations},
(Cambridge: Cambridge University Press)
\bibitem[Alexander et al. 2006]{alex06}Alexander, R. D., Clarke, C. J., Pringle, J. E., 2006, MNRAS, 369, 229
\bibitem[Alibert et al. 2005]{ali05} Alibert Y. , Mordasini C., Benz W. ,  Winisdoerffer C., 2005, A\&A 434, 343
\bibitem[Balbus  \& Hawley 1991]{BH91}Balbus, S. A. , Hawley, J. F., 1991, ApJ, 376, 214-222
\bibitem[Butler et al. 2004]{But04}Butler, R.~P.,
Vogt, S.~S., Marcy, G.~W., et al., 2004, ApJ, 617, 580
\bibitem[Coughlin et al. 2008]{cou08} Coughlin, J.~L., Stringfellow, G.~S., Becker, A.~C.,
et al.,  2008, arXiv:0809.1664
\bibitem[Fogg \& Nelson 2005]{FN05}Fogg, M. J.,  Nelson, R. P., 2005,  A\&A,  441, 791
\bibitem[Fogg \& Nelson 2007]{FN07}Fogg M.J., Nelson R.P, 2007, A\&A 472,1003
   \bibitem[Gammie 1996]{gam96}Gammie C.F., 1996, ApJ 457,355
\bibitem[Goldreich \& Tremaine 1979]{GT79}Goldreich, P., Tremaine,S., 1979, ApJ, 233,857
\bibitem[Haicsh et al. 2001]{Hai01}Haicsh,K.E.,Jr., Lada,E.A.,  Lada,C.J., 2001, ApJ,241, 425
 \bibitem[Hayashi 1981]{hay81}Hayashi,C. 1981,  Prog. Theor.Phys. Suppl., 70,35
  \bibitem[Ida \& Lin 2004]{idalin04}Ida, S., Lin, D. N. C., 2004, ApJ,  604, 388
\bibitem[Ida \& Lin 2008]{idalin08} Ida, S.,  Lin, D. N. C., 2008,  ApJ, 673, 487
\bibitem[Kokubo \& Ida 1996]{KI96} Kokubo, E., Ida, S. 1996, Icarus, 123, 180
\bibitem[Kokubo \& Ida 2002]{KI02} Kokubo, E., Ida, S. 2002, ApJ, 581,666
   \bibitem[K\"{o}nigl 1991]{Kon91}K\"{o}nigl,A., 1991, ApJ, 370,L39
\bibitem[Kretke \& Lin 2007]{KL07}  Kretke,K. A., Lin,D.N.C., 2007, ApJ, 664, L55
\bibitem[Kretke et al. 2009]{KL09} Kretke, K. A., Lin, D. N. C., Garaud,P.,  Turner,N. J.,  2009,  ApJ,  690, 407.
\bibitem[Lin et al. 2009]{lin09}Lin, D.N.C., Zhou, J.L., Wang, S., Kretke, K. A., in preparation.
\bibitem[Maness et al. 2007]{man07}Maness, H.~L., Marcy, G.~W., Ford, E.~B., et al., 2007, PASP, 119, 90
\bibitem[Mardling 2007]{mar07}Mardling, R.~A.,  2007, MNRAS, 382, 1768
\bibitem[Mardling \& Lin 2004]{ML04}Mardling, R.,  Lin, D. N.C. 2004, ApJ,614,955-959
\bibitem[Masset 2002]{masset02}Masset, F.S. 2002 A\&A,387,605
\bibitem[Masset et al. 2006]{mas06}Masset, F. S., Morbidelli, A., Crida, A., Ferreira, J. 2006,ApJ,642, 478
\bibitem[Mayor et al. 2008]{Mayor08} Mayor,M., Udry,S., Lovis ,C. et al., 2009,
 A \& A, 493, 639
\bibitem[Morbidelli et al. 2008]{Mor08} Morbidelli A.,  Crida, A., Masset, F., Nelson, R. P.,  2008, A\&A 478, 929
\bibitem[Nagasawa et al. 2005]{nag05}Nagasawa, M., Lin, D. N. C.,  Thommes, E. 2005, ApJ, 635,578
\bibitem[Pollack et al. 1996]{pol96}Pollack, J. B., Hubickyj, O., Bondenheimer, P., Lissauer, J. J., Podolack,
 M. \& Greenzweig, Y. 1996, Icarus, 124, 62-85,
 \bibitem[Raymond et al. 2008]{ray08}Raymond, S. N.,  Barnes, R.,  Mandell, A. M., 2008, MNRAS, 384, 663
 \bibitem[Ribas et al. 2008]{rib08a} Ribas, I., Font-Ribera, A.,  Beaulieu, J.-P.,  2008, ApJL, 677, L59
 \bibitem[Safronov 1969]{Saf69}Safronov, V.S. 1969, {\it Evolution of the Protoplanetary
    Cloud and Formation of the Earth and the planets}, English translation NSSA TT F-677(1972)
\bibitem[Shakura \& Sunyaev 1973]{SS73}Shakura, N. I. , Sunyaev, R.A., 1973, A\&A 24, 337
\bibitem[Tanaka et al. 2002]{Tan02}Tanaka, H.,Takeuchi,T.,  Ward,W.R., 2002,  ApJ, 565, 1257
\bibitem[Terquem  \& Papaloizhou 2007]{tp07}Terquem,C.,  Papaloizhou,J.B.C., 2007, ApJ,654,1110
\bibitem[Tong \& Zhou 2009]{TZ09}Tong,X. \& Zhou,J.L., 2009, accepted by Science in China (G): Physics, Mechanics and Astronomy. arXiv 0812.3195.
\bibitem[Ward 1991]{ward91}Ward, W.R., 1991, Lunar Planet Sci. Conf. 22, 1463
\bibitem[Ward 1993]{ward93}Ward, W.R., 1993, Icarus, 106, 274
\bibitem[Ward 1997]{Ward97}Ward, W.R., 1997, Icarus, 126, 261
\bibitem[Zhou et al. 2005]{zhou05}Zhou, J.L., Aarseth, S. J., Lin, D. N. C., Nagasawa, M., ApJ, 631, L85
\bibitem[Zhou et al. 2007]{zls07}Zhou, J.L., Lin, D.N.C., Sun,Y.S, 2007, ApJ, 666, 423
\bibitem[Zhou \& Lin 2008]{ZL08}Zhou, J.L. ,  Lin, D.N.C. 2008 in {\it
Exoplanets: Detection, Formation and Dynamics}, eds: Sun,Y.S.,
Ferraz-Mello,S., Zhou,J. L., Proc. of IAU Symp. 249, Cambridge,
Cambridge unversity Press, 2008, 285

\end{thebibliography}
\end{document}